\documentclass{iaus}
\usepackage{graphicx}
\usepackage{caption}

\title[Massive eclipsing binaries in Westerlund 1] 
{Fundamental Parameters of 4 Massive Eclipsing Binaries in Westerlund 1}

\author[Koumpia \& Bonanos]   
{E. Koumpia
 \and A.Z. Bonanos\footnote{EK \& AZB acknowledge research and travel support from the IAU and the European Commission 
Framework Program Seven under the Marie Curie International Reintegration Grant 
PIRG04-GA-2008-239335.}}

\affiliation{National Observatory of Athens, Institute of Astronomy \& Astrophysics, \\ I. Metaxa \& Vas. Pavlou St., 
Palaia Penteli GR-15236 Athens, Greece \\[\affilskip]
{\tt koumpia@astro.noa.gr}, {\tt bonanos@astro.noa.gr} \\[\affilskip]}

\pubyear{2011}
\volume{282}  
\pagerange{1--2}
\jname{From Interacting Binaries to Exoplanets:\\Essential Modelling Tools}
\editors{A.C. Editor, B.D. Editor \& C.E. Editor, eds.}
\begin{document}

\maketitle

\begin{abstract}

We present fundamental parameters of 4 massive eclipsing binaries in the young 
massive cluster Westerlund 1. The goal is to measure accurate masses and radii 
of their component stars, which provide much needed constraints for
evolutionary 
models of massive stars. Accurate parameters can further be used to determine a 
dynamical lower limit for the magnetar progenitor and to obtain an independent 
distance to the cluster. Our results confirm and extend the evidence for a high 
mass for the progenitor of the magnetar.
 
\end{abstract}

\textbf{1. Introduction}

Westerlund 1 is one of the most massive young clusters known in the Local Group of galaxies, with an age of 3-5 Myr. It 
was discovered by
\cite[Westerlund (1961)]{Westerlund61},
but remained unstudied until recently due to the high interstellar extinction in its direction. It contains an assortment 
of rare evolved high-mass stars, such as blue, yellow and red supergiants, Wolf-Rayet stars, a luminous blue variable, 
many OB supergiants, as well as 4 massive eclipsing binary systems (Wddeb, Wd13, Wd36, WR77o, see
\cite[Bonanos 2007]{Bonanos07}).
Furthermore, the magnetar CXO J164710.2-455216 was discovered in the cluster by \cite[Muno et al. (2006)]{Muno_etal06} in X-rays. 
This magnetar is a slow X-ray pulsar that is assumed to have formed from a massive progenitor star.

Eclipsing binaries provide the only accurate way for the measurement of masses and radii of stars. Thus, the study of 
these systems in the cluster is important for the following reasons: (1) the determination of fundamental parameters of 
the component stars (mass, radii, etc.), in order to increase the small sample of massive stars with well known physical 
parameters, (2) the test of stellar models for the formation and the evolution of massive stars, 
(3) the determination of a dynamical lower limit for the mass of the magnetar progenitor and (4) EBs present a great 
opportunity for an independent measurement of the distance, based on the expected absolute magnitude. 

\textbf{2. The individual binaries}

We have analyzed spectra of all 4 eclipsing binaries, taken in 2007-2008 with the 6.5 meter Magellan telescope at 
Las Campanas Observatory, Chile. The spectra were reduced and extracted using IRAF\footnote[2]{IRAF is distributed 
by the NOAO, which are operated by the Association of Universities for Research in Astronomy, Inc., under cooperative 
agreement with the NSF.}. For the determination of the radial velocities we adopted a $\chi^2$ minimization technique, 
which finds the least $\chi^2$ from the observed spectrum and fixed synthetic TLUSTY models 
(\cite[Lanz \& Hubeny 2003]{LanzHubeny03}).
For this purpose we used the narrow Helium lines ($\lambda\lambda6678,7065$), as they are less sensitive 
to systematics, rather than the broader hydrogen lines. 

The physical and orbital parameters of the members of the 4 systems (period, masses, radii, surface gravities of 
the components, eccentricity, inclination etc.), resulted from modelling light and radial velocity curves using PHOEBE 
software
(\cite[Prsa \& Zwitter 2005]{PrsaZwitter05}). Our final results will be presented in Koumpia \& Bonanos (in prep.).

\textbf{\underline{Wd13}}: is a semi-detached, double-lined spectroscopic binary (B0.5Ia$^{+}$/WNVL and O9.5-B0.5I types,
\cite[Ritchie et al. 2010]{Ritchie_etal10}) in a circular orbit (see Fig.\,\ref{fig1}). Its spectra show both 
absorption and emission lines. It provides the first dynamical constraint on the mass of the magnetar progenitor and has important implications for the threshold mass which gives rise to black holes versus neutron stars.\\
\textbf{\underline{Wd36}}: is an overcontact, double-lined spectroscopic binary system (OB-type) in a circular orbit. 
\\
\textbf{\underline{WR77o}}: is probably a double contact binary system in an almost circular orbit. It is a single lined 
spectroscopic binary. The spectroscopic 
visible star is a Wolf-Rayet star of spectral type WN6-7 
(\cite[Negueruela \& Clark 2005]{NegueruelaClark05}), 
with line widths of 2000 km~s$^{-1}$.\\
\textbf{\underline{Wddeb}}: is a double-lined spectroscopic binary system (OB-type) with an eccentricity of 
almost 0.2. Being a detached system, it contains the first main-sequence stars detected in 
the cluster, yielding masses of 2 of the most massive unevolved stars in Wd1.

\textbf{3. Conclusions}

Having obtained the physical and orbital parameters of each system, we confirm the high mass of 
Wd13 and therefore the high mass of the magnetar progenitor, also found by the independent study 
of \cite[Ritchie et al. (2010)]{Ritchie_etal10}. 
We also compared our results with stellar models of evolution of single stars (\cite[Claret 2004]{Claret04}). 

\begin{center}
	\includegraphics[scale=0.23]{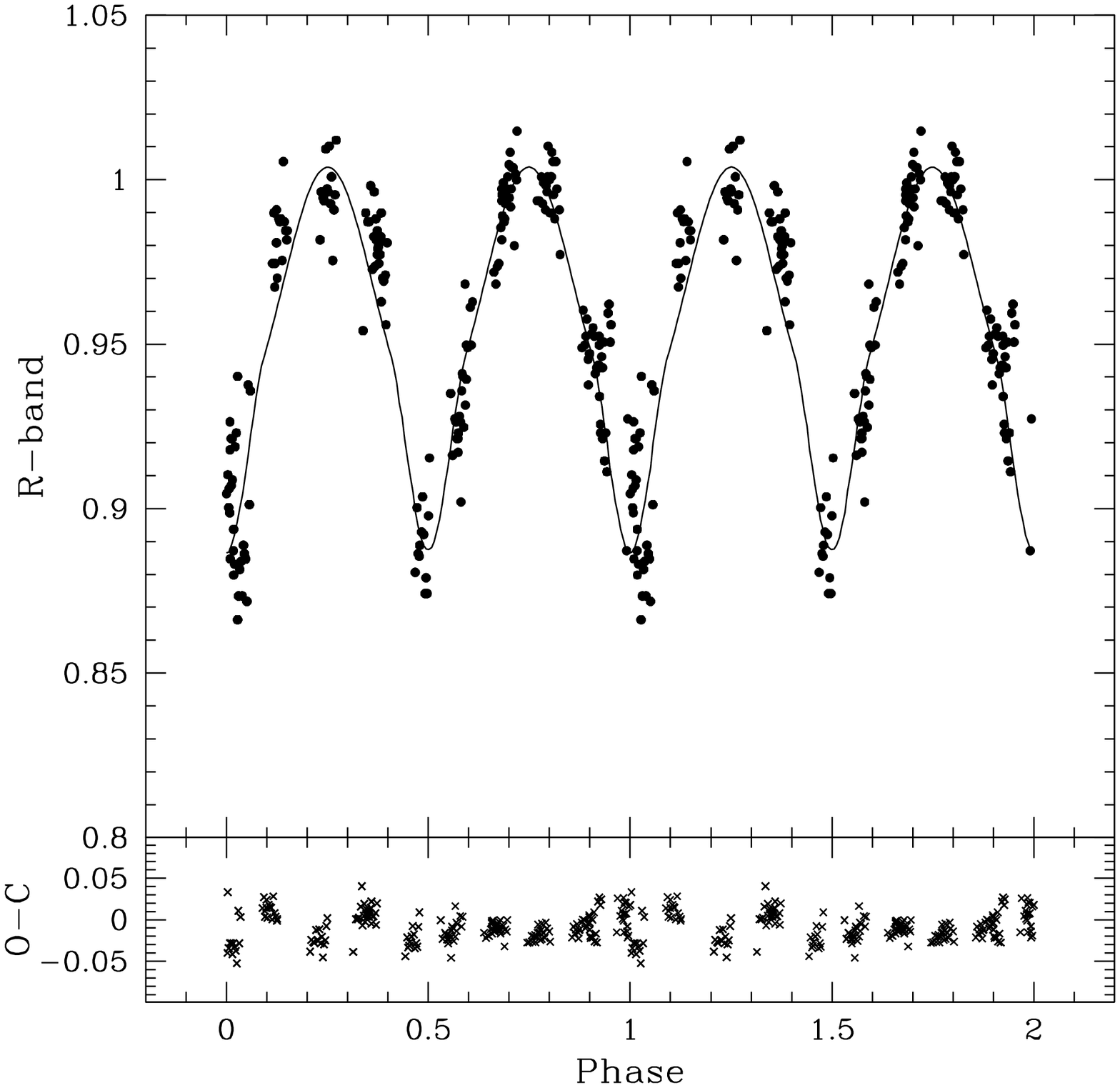}
\includegraphics[scale=0.23]{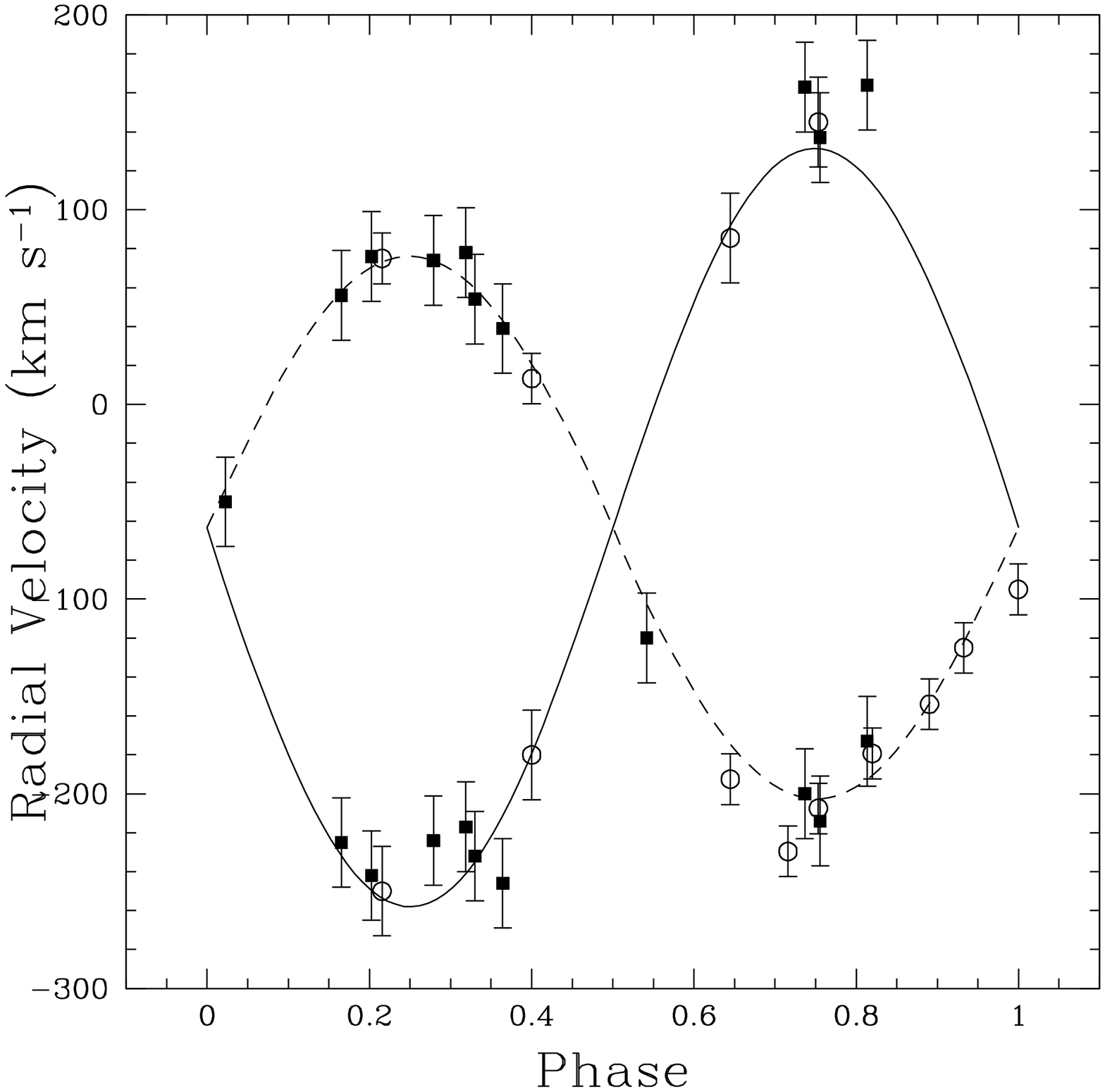}
	\firstsection
 \captionof{figure}{Phased $R$ light curve and radial velocity curve of Wd13, respectively. Radial velocity points 
from \cite[Ritchie et al. (2010)]{Ritchie_etal10} are also included (black squares).}
   	\label{fig1}
\end{center}

\firstsection

\end{document}